\documentclass[conference]{IEEEtran}
\IEEEoverridecommandlockouts

\usepackage{cite}
\usepackage{amsmath,amssymb,amsfonts}
\usepackage{algorithmic}
\usepackage{graphicx}
\usepackage{textcomp}
\usepackage{xcolor}
\usepackage{tikz}

\usepackage{amsthm}

\newcommand{\bs}[1]{\boldsymbol{#1}}
\newcommand{\bm}[1]{\mathbf{#1}}

\begin{document}

\title{Consensus-Based Distributed Filtering for Angles-Only Tracking in Cislunar Space\\

\thanks{This work was partially supported by the Italian Ministry of Education and Research (MUR) within the framework of the FoReLab project (Departments of Excellence), by the Italian Space Agency (Agenzia Spaziale Italiana, ASI) under the research project "DORA - Determinazione Orbitale Relativa Autonoma on-board, con applicazioni terrestri e lunari," through agreement no. 2024-35-HH.0 (CUP n. F53C24000360001) between ASI and the University of Pisa.}
\thanks{Accepted for presentation at IEEE NAVICON 2026, Munich, Germany,
December 9–11, 2026.}
\thanks{© 2026 IEEE. Personal use of this material is permitted. Permission from
IEEE must be obtained for all other uses, in any current or future media,
including reprinting/republishing this material for advertising or promotional
purposes, creating new collective works, for resale or redistribution to servers
or lists, or reuse of any copyrighted component of this work in other works.}
}

\author{\IEEEauthorblockN{Giovanni Romagnoli}
\IEEEauthorblockA{\textit{Department of Information Engineering} \\
\textit{University of Pisa}\\
Pisa 56122, Italy \\
giovanni.romagnoli@phd.unipi.it}
\and
\IEEEauthorblockN{Fabrizio Pilone}
\IEEEauthorblockA{\textit{Department of Information Engineering} \\
\textit{University of Pisa}\\
Pisa 56122, Italy \\
fabrizio.pilone@phd.unipi.it}
\and
\IEEEauthorblockN{Giordana Bucchioni}
\IEEEauthorblockA{\textit{Department of Information Engineering} \\
\textit{University of Pisa}\\
Pisa 56122, Italy \\
giordana.bucchioni@unipi.it}
}

\maketitle

\begin{abstract}

This paper investigates consensus-based distributed filtering strategies for multi-observer angles-only navigation in the cislunar space. The objective is to estimate the inertial state of a target spacecraft using only line-of-sight measurements collected by a network of observer spacecraft. This problem is particularly relevant for autonomous cislunar navigation, where ground support may be limited, and centralized fusion architectures may be problematic due to communication and computational constraints, and single-point failure risks. Three distributed filtering approaches are considered: the Distributed Extended Kalman Filter, the Information-Weighted Consensus Filter, and the Kalman Consensus Filter. All methods share the same local prediction and measurement-processing steps, but differ in the way information is exchanged and fused across the network. Their performance is evaluated and compared to a centralized Extended Kalman Filter, adopted as benchmark, and independent local filters in a high-fidelity cislunar scenario. Numerical simulations show that the distributed approaches significantly improve estimation accuracy with respect to independent filters. Monte Carlo sensitivity analyses are then conducted to characterize the robustness of the considered methods to finite consensus iterations and prior misalignment among the agents. 
\end{abstract}

\section{Introduction}
The successful return of astronauts to the lunar vicinity represents only the beginning of a sustained human and robotic presence in the cislunar domain, namely the region of space between the Earth and the Moon. In the next decade, more than 30 missions are expected to target the Earth-Moon region with both scientific and commercial applications \cite{baker2024comprehensive}.

In this context, autonomous on-board navigation is a critical capability for safe and efficient operations \cite{pilone2025relative}. Indeed, communication with ground stations is often limited by visibility windows and signal delays, making continuous ground-based navigation support impractical. Among the available sensing technologies, passive optical sensors represent a preferred choice due to their low weight and power requirements, as well as an extended dynamic range and high accuracies \cite{palo2013agile}. These sensors measure the line-of-sight (LOS) vector that corresponds to the direction between the sensor and the observed target, but they do not provide direct range information. As a consequence, a single observer is fundamentally limited in its ability to reconstruct the true target state due to the well known observability problem \cite{woffinden2009observability}.
This limitation is especially relevant in the cislunar environment, where the highly non-linear dynamics further complicate single-sensor navigation. 

This motivates the adoption of multi-agent navigation approaches. By fusing LOS measurements from spatially distributed spacecraft, geometric ambiguities can be resolved, and the local measurement information can be shared across the network to drastically improve tracking accuracy. As a consequence, in the last few years, a series of distributed angles-only architectures for cooperative tracking have been proposed in both Earth-orbiting and lunar/cislunar scenarios  \cite{hu2021distributed, iiyama2022autonomous
, jia2016cooperative}. Consensus-based distributed filters are particularly well suited for cooperative tracking in satellite networks, as they distribute the estimation burden across the agents, remove the need for a central fusion node, and avoid single-point failures. Moreover, they require only local information exchange among neighboring spacecraft, making them scalable and robust to communication constraints. Several approaches have been proposed, including the Distributed Extended Kalman Filter (DEKF)~\cite{long2012distributed}, the Information-Weighted Consensus Filter (ICF)~\cite{kamal2012information}, and the Kalman Consensus Filter (KCF)~\cite{olfati2009kalman}. 
Under ideal conditions, with unlimited communication, and aligned prior estimates across the network, these methods can approach the performance of a centralized estimator, i.e., a filter that collects and processes all measurements at a single fusion node.
In operational scenarios, however, communication is finite and agents may
be initialized with different local estimates, so that the performance of
distributed filters can strongly depend on the communication budget and on
the degree of prior mismatch.

This work investigates these aspects by implementing and comparing three consensus-based filtering strategies, namely the DEKF, the ICF, and the KCF, in a cislunar angles-only tracking scenario, where a formation of observer spacecraft on a Near-Rectilinear Halo Orbit estimates the state of a target using only noisy LOS measurements. A centralized EKF serves as benchmark, while independent local filters are included to quantify the benefit of inter-agent information exchange. Monte Carlo sensitivity analyses with respect to finite consensus iterations and prior misalignment are conducted to characterize both the accuracy and the robustness of the considered methods under realistic constraints.

The remainder of the paper is organized as follows. Section~\ref{sec: dynamics_and_meas} introduces the high-fidelity dynamical model used for simulation, the simplified on-board filter model, and the LOS measurement model. Section~\ref{sec:filters} describes the considered distributed filtering strategies and the centralized reference estimator. Section~\ref{results} presents the proposed simulation scenario and the numerical simulation results. Finally, Section~\ref{sec:conc} concludes the work.

\section{Dynamics and Measurement Model}
\label{sec: dynamics_and_meas}
\subsection{High-Fidelity Ephemeris Model}
\label{sec:full_dyn}
The reference trajectories of the spacecraft are propagated in the Earth-Centered Inertial (ECI) frame using planetary ephemerides from the SPICE dataset \cite{acton2016spice}. The acceleration of each spacecraft explicitly accounts for the gravitational attraction of the Earth and the Moon, and is given by:
\begin{equation}
\label{eq:eph acc}
\bs{\ddot{r}}= -\mu_{E}\frac{\bs{r}}{\|{\bs{r}\|}^3}-\mu_{M}\left(\frac{\bs{r}-\bs{r}_M}{\|{\bs{r}-\bs{r}_M\|}^3}+\frac{\bs{r}_{M}}{\|{\bs{r}_{M}\|}^3}\right) + \bs{d}
\end{equation}

where $\bs{r}$ and $\bs{r}_M$ are the ECI positions of the spacecraft and Moon, while $\mu_{E}$ and $\mu_{M}$ are the gravitational parameters of the Earth and Moon respectively. The term $\bs{d}$ is the deterministic solar third-body acceleration, computed from ephemerides and included only in the reference propagation.
The positions of the three bodies are obtained from the ephemeris catalog at every integration step. 

\subsection{Simplified Filter Model}
\label{sec:simp_dyn}
For on-board applications, the full ephemeris propagation is computationally expensive and relies on real-time access to planetary position data. For this reason, a simplified dynamical model is used for on-board filtering. In particular, the dominant gravitational effects are maintained, while the Moon's trajectory is analytically approximated. 

Following the assumptions of the planar Circular Restricted Three-Body Problem (CR3BP) \cite{vallado2001fundamentals}, the Earth and the Moon are assumed to move on circular coplanar orbits about their common center of mass. Equivalently, when expressed in the geocentric ECI frame, the Moon is modeled as moving on a circular orbit around the Earth, with the orbital plane  and angular velocity fixed to their values at the initial epoch $t_0$. 
Let $\bs{r}_{M0}$ and $\bs{v}_{M0}$ denote the initial position and velocity of the Moon in the ECI frame. The initial orbital angular momentum is then defined as $\bs{h}_{M0}=\bs{r}_{M0}\times\bs{v}_{M0}$, and the unit vectors spanning the orbital plane are introduced as
\begin{equation}
\label{eq:orbital plane unit vectors}
\hat{\bs{u}}_1 = \frac{\bs{r}_{M0}}{\|\bs{r}_{M0}\|}, 
\qquad
\hat{\bs{u}}_2 = \frac{\bs{h}_{M0}\times\bs{r}_{M0}}{\|\bs{h}_{M0}\times\bs{r}_{M0}\|}.
\end{equation}
Assuming uniform circular motion, the mean angular velocity can be computed as $\omega=\|\bs{h}_{M0}\|/\|\bs{r}_{M0}\|^2$, and the lunar position at elapsed time $t$ from $t_0$ can be approximated as
\begin{equation}
\label{eq:approx. Moon pos}
\bs{r}_M(t)=\|\bs{r}_{M0}\|\left(\cos(\omega t)\hat{\bs{u}}_1 + \sin(\omega t)\hat{\bs{u}}_2\right).
\end{equation}
Finally, the actual equation of motion implemented in the filtering algorithm is obtained by substituting \eqref{eq:approx. Moon pos} in \eqref{eq:eph acc} and neglecting the disturbance term. 

The resulting dynamical mismatch reflects the process noise that a real onboard filter must absorb, and is handled by the process noise covariance $\bm{Q}$ in the filter.

\subsection{Measurement Model}
\label{sec:noise}
Each observer satellite $i$ is equipped with a passive optical sensor that provides a noisy line-of-sight (LOS) unit vector to the target. Let
\begin{equation}
\label{eq:true_los}
\bs{\ell}_i^{true} = h_i(\bs{x}) = \frac{\bs{r}_t-\bs{r}_i}{\|\bs{r}_t-\bs{r}_i\|},
\end{equation}
where $\bs{x}=[\bs{r}_t^\top,\bs{v}_t^\top]^\top \in \mathbb{R}^{6}$ is the target state. The measured LOS is simulated as a small random rotation of the true vector:
\begin{equation}
\label{eq:noisy_los}
\bs{\ell}_i = \exp([\bs{\theta}_i]_\times)\bs{\ell}_i^{true},
\end{equation}
where $\bs{\theta}_i \in \mathbb{R}^3$ is a zero-mean isotropic Gaussian rotation vector, $\bs{\theta}_i\sim\mathcal{N}(\bs{0},\sigma_\theta^2\bm{I}_3)$, and $[\cdot]_\times$ denotes the skew-symmetric cross-product operator. All sensors share the same $\sigma_\theta$, with independent noise realizations across sensors and epochs.
For small angles, this model can be approximated as
\begin{equation}
\label{eq:noisy lin los}
\bs{\ell}_i \simeq h_i(\bs{x})+\bs{\nu}_i, \qquad
\bs{\nu}_i \sim \mathcal{N}(\bs{0},\bm{R}_i),
\end{equation}
where the covariance matrix
\begin{equation}
\label{eq: R matrix}
\bm{R}_i = \sigma_\theta^2 \left( \bm{I}_3 -\bs{\ell}_i\bs{\ell}_i^\top \right) \in \mathbb{R}^{3\times3}
\end{equation}
is used as the measurement weighting matrix in the filtering update.

\section{Navigation Algorithms}
\label{sec:filters}
Let $\bs{x}_k$ be the target state at time $t_k$. Each agent $i$ maintains a local estimate of the state $\hat{\bs{x}}_k^i$ and the associated covariance matrix $\bm{P}_k^i \in \mathbb{R}^{6\times6}$. The communication network is described by an undirected graph $\mathcal{G}=(\mathcal{V},\mathcal{E})$, where $\mathcal{V}$ is the set of observers, $\mathcal{E}$ contains the pairs of nodes that can directly communicate, and $\mathcal{N}_i$ denotes the set of neighbors of agent $i$. Therefore, the total number of agents is the cardinality of $\mathcal{V}$ and is denoted as $N=|\mathcal{V}|$.

\subsection{Shared Prediction Step}
\label{sec:predict_step}
At each navigation step, each observer locally propagates its current estimate using the simplified dynamical model introduced in the previous section. The prediction step can be formulated as
\begin{equation}
\label{prediction}
\begin{gathered}
\hat{\bs{x}}_{k|k-1}^i=f(\hat{\bs{x}}_{k-1}^i),\\
\bm{P}_{k|k-1}^i=\bm{F}_{k-1}^i\bm{P}_{k-1}^i(\bm{F}_{k-1}^i)^\top+\bm{Q},
\end{gathered}
\end{equation}
where $\bm{Q}\in\mathbb{R}^{6\times6}$ is the process noise covariance that accounts for unmodeled contributions in the discretized, with step $\Delta t$, state prediction function $f(\cdot)$, while $\bm{F}_{k-1}^i$ is the process Jacobian matrix
\begin{equation}
\label{eq:process_jacobian}
\bm{F}_{k-1}^i=\left.\frac{\partial f}{\partial \bs{x}}\right|_{\hat{\bs{x}}^i_{k-1}}=
\begin{bmatrix}
\bm{I}_3 & \Delta t\,\bm{I}_3\\
\Delta t\,\bm{G}_{k-1}^i & \bm{I}_3
\end{bmatrix}\in\mathbb{R}^{6\times6}.
\end{equation}
The matrix $\bm{G}_{k-1}^i$ is the gravity-gradient contribution evaluated by agent $i$ at the current state estimate: 
\begin{equation}
\label{eq:G_matrix}
\bm{G}_{k-1}^i=-\mu_E\bm{\Gamma}(\hat{\bs{r}}^i_{t,k-1})-\mu_M\bm{\Gamma}(\hat{\bs{r}}^i_{t,k-1}-\bs{r}_{M,k-1}),
\end{equation}
where the operator $\Gamma$ is defined as:
\begin{equation}
\label{eq:gamma_operator}
\bm{\Gamma}(\bs{a})=\frac{\bm{I}_3}{\|\bs{a}\|^3}-3\frac{\bs{a}\bs{a}^\top}{\|\bs{a}\|^5}.
\end{equation}

\subsection{Local Information Terms Computation}
At each navigation step $k$, observer $i$ acquires a LOS measurement $\bs{\ell}^i_k$ and linearizes the measurement function around the predicted state. The measurement Jacobian $\bm{H}_k^i\in\mathbb{R}^{3\times6}$ is
\begin{equation}
\label{eq:meas_jacobian}
\bm{H}^i_k=\left.\frac{\partial h_i}{\partial \bs{x}}\right|_{\hat{\bs{x}}^i_{k|k-1}}=
\begin{bmatrix}
\bm{H}^i_{r,k} & \bm{0}_{3\times3}
\end{bmatrix},
\end{equation}
where
\begin{equation}
\label{eq:H_r_block}
\bm{H}^i_{r,k}=\frac{1}{\|\bs{\rho}^i_k\|}\left(\bm{I}_3-\frac{\bs{\rho}^i_k(\bs{\rho}^i_k)^\top}{\|\bs{\rho}^i_k\|^2}\right),
\end{equation}
and $\bs{\rho}^i_k=\hat{\bs{r}}^i_{t,k|k-1}-\bs{r}^i_k
$ is the predicted relative position between the target and agent $i$ at time $t_k$.

Each agent then locally computes the following information quantities:
\begin{align}
\label{eq:u}
\bs{u}^i_k &= (\bm{H}^i_k)^\top (\bm{R}_k^i)^{-1}
\bs{\ell}^i_k,\\
\label{eq:uhat}
\hat{\bs{u}}^i_k &= (\bm{H}^i_k)^\top (\bm{R}_k^i)^{-1}
h_i(\hat{\bs{x}}^i_{k|k-1}),\\
\label{eq:U}
\bm{U}^i_k &= (\bm{H}^i_k)^\top (\bm{R}_k^i)^{-1}
\bm{H}^i_k \in \mathbb{R}^{6\times6}.
\end{align}
Here $\bs{u}^i_k\in \mathbb{R}^6$ and $\hat{\bs{u}}^i_k \in \mathbb{R}^6$
are the information-weighted versions of the actual and
predicted LOS, while $\bm{U}^i_k$ quantifies how informative agent $i$'s measurement is about the target state. $\bm{R}_i$ denotes the ideal LOS covariance associated with the optical
measurement model introduced in~\eqref{eq: R matrix}. At each epoch $t_k$,
this covariance is evaluated using the LOS measurement collected by agent
$i$, leading to the time-dependent matrix $\bm{R}_k^i$ used in the filter
update. Since the optical measurement provides angular information only,
the ideal covariance is singular by construction, because its nonzero
variance is confined to the plane orthogonal to the measured LOS. A small
regularization term is therefore added along the LOS direction for
numerical stability, with negligible effect on the angular information
content:
\begin{equation}
\label{eq:R_reg}
\bm{R}_k^i = \sigma_\theta^2 \left[
\left( \bm{I}_3 -\bs{\ell}_k^i\bs{\ell}_k^{i\top} \right)
+ \epsilon_{\mathrm{reg}}\bs{\ell}_k^i\bs{\ell}_k^{i\top}
\right].
\end{equation}
Given \eqref{eq:u}-\eqref{eq:U}, the innovation information vector can also
be defined as
\begin{equation}
\label{eq:innov_info}
\bs{\iota}^i_k =
(\bs{u}^i_k - \hat{\bs{u}}^i_k)
+ \bm{U}^i_k\hat{\bs{x}}^i_{k|k-1}.
\end{equation}

\subsection{Average Consensus Protocol}
The distributed fusion step leverages an average consensus protocol \cite{xiao2006distributed} over the communication graph $\mathcal{G}=(\mathcal{V},\mathcal{E})$ in order to drive locally computed quantities toward their average across the network. 

Let $\xi$ be a generic variable. Starting from the local values $\xi_0^i$, each agent performs $N_c$ consensus iterations. At iteration $q$, the update is given by:
\begin{equation}
\label{eq:consensus}
\xi_{q+1}^{i}=w_{ii}\xi_q^{i}+\sum_{j\in\mathcal{N}_i}w_{ij}\xi_q^{j}, \quad q=0,\ldots,N_c-1.
\end{equation}
The Metropolis weights are $w_{ij}=1/[1+\max(d_i,d_j)]$ for $j\in\mathcal{N}_i$, $w_{ii}=1-\sum_{j\in\mathcal{N}_i}w_{ij}$, and zero otherwise, where $d_i=|\mathcal{N}_i|$ \cite{xiao2006distributed}. 
For an undirected connected graph, these weights preserve the network average and guarantee convergence to it as $N_c\to\infty$. Each navigation step uses only $N_c$ iterations, yielding a finite-consensus approximation.

\subsection{Distributed Filtering Strategies}

The three filters share the same prediction step and the same local
measurement-information computation. The DEKF and ICF apply average consensus after computing the local information terms and before the posterior update, to approximate centralized information fusion, while the KCF keeps the update local to the one-hop
neighborhood and adds an explicit correction on the disagreement between
neighboring predicted estimates.

\subsubsection{Distributed Extended Kalman Filter (DEKF)}
In the Distributed Extended Kalman Filter (DEKF) formulation, the consensus protocol is applied to local measurement information terms $\bs{u}^i_k$, $\hat{\bs{u}}^i_k$, and $\bm{U}^i_k$. After $N_c$ consensus iterations, agent $i$ ends up with local information terms $\bar{\bs{u}}^i_k$, $\bar{\hat{\bs{u}}}^i_k$, and $\bar{\bm{U}}^i_k$.
The local update step is then computed as: 
\begin{equation}
\begin{gathered}
\label{eq:dekf_upd}
\bm{P}^i_k=\left[(\bm{P}^i_{k|k-1})^{-1}+\bar{\bm{U}}^i_k\right]^{-1},\\
\hat{\bs{x}}^i_k=\hat{\bs{x}}^i_{k|k-1}+\bm{P}^i_k\left(\bar{\bs{u}}^i_k-\bar{\hat{\bs{u}}}^i_k\right).
\end{gathered}
\end{equation}
In this formulation, $\bm{P}_k^i$ is interpreted as a local-filter covariance scaled with respect to the centralized covariance, i.e., $\bm{P}^i_k \simeq N\bm{P}^c_k$ under ideal consensus conditions. Therefore, in the DEKF implementation only, the prediction step uses the scaled process noise $N\bm{Q}$ to preserve this covariance scaling.
It has been demonstrated that, under the hypotheses of aligned prior estimates among agents and $N_c \to \infty$, the local filters converge to centralized performance \cite{long2012distributed}. 

\subsubsection{Information-Weighted Consensus Filter (ICF)}
The Information-Weighted Consensus Filter (ICF) \cite{kamal2012information} applies consensus to quantities that include both the prior information and the local measurement contribution. Let
\begin{equation}
\label{eq:prior_information}
\bm{Y}^i_{k|k-1}=(\bm{P}^i_{k|k-1})^{-1}, \qquad \bs{y}^i_{k|k-1}=\bm{Y}^i_{k|k-1}\hat{\bs{x}}^i_{k|k-1}
\end{equation}
be the predicted information matrix and information vector of agent $i$. Each agent computes 
\begin{equation}
\label{eq:icf_terms}
\bm{V}^i_k=\frac{1}{N}\bm{Y}^i_{k|k-1}+\bm{U}^i_k,\qquad
\bs{v}^i_k=\frac{1}{N}\bs{y}^i_{k|k-1}+\bs{\iota}^i_k.
\end{equation}
Consensus is then performed on the pair $(\bm{V}^i_k,\bs{v}^i_k)$. Denoting the corresponding consensus outputs as $\bar{\bm{V}}^i_k$ and $\bar{\bs{v}}^i_k$, the posterior information matrix and vector are reconstructed as
\begin{equation}
\label{eq:icf_inf_upd}
\bm{Y}^i_k=N\bar{\bm{V}}^i_k, \qquad \bs{y}^i_k=N\bar{\bs{v}}^i_k.
\end{equation}
The posterior state and covariance estimates are finally computed as
\begin{equation}
\label{eq:icf_state_upd}
\bm{P}^i_k=(\bm{Y}^i_k)^{-1}, \qquad \hat{\bs{x}}^i_k=\bm{P}^i_k\bs{y}^i_k.
\end{equation}
Similarly to the DEKF, convergence to the centralized information update is guaranteed under ideal consensus, i.e., $N_c\to\infty$, and consistent prior information among the agents \cite{kamal2012information}. However, differently from the DEKF, the ICF includes the local prior information directly into the consensus variables. As a result, the ICF is generally more robust to finite consensus iterations and small prior misalignments.

\subsubsection{Kalman Consensus Filter (KCF)}

The Kalman Consensus Filter (KCF) \cite{olfati2009kalman} combines a neighborhood-based measurement update with an explicit consensus correction on the predicted state estimates. Differently from the DEKF and ICF, no iterative average-consensus reconstruction is performed in this formulation. Instead, each agent performs a cooperative update using the measurement information available within its one-hop neighborhood and a state-consensus correction term. Let $\mathcal{S}_i=\mathcal{N}_i\cup\{i\}$
be the set containing agent $i$ and its neighbors. The one-hop fused information matrix and vector are defined as
\begin{equation}
\label{eq:one_step_fus}
\bm{U}^{i,\mathcal{S}}_k=\sum_{j\in\mathcal{S}_i}\bm{U}^j_k,\qquad
\bs{\iota}^{i,\mathcal{S}}_k=\sum_{j\in\mathcal{S}_i}\bs{\iota}^j_k.
\end{equation}
The posterior covariance of agent $i$ is computed from the neighborhood information as
\begin{equation}
\label{eq:kcf_cov}
\bm{P}^{i}_k=\left[(\bm{P}^i_{k|k-1})^{-1}+\bm{U}^{i,\mathcal{S}}_k\right]^{-1}.
\end{equation}
The corresponding measurement-related correction is
\begin{equation}
\label{eq:kcf_meas_corr}
\Delta\hat{\bs{x}}^{i,\mathrm{meas}}_k = \bm{P}^{i}_k \left(
\bs{\iota}^{i,\mathcal{S}}_k -
\bm{U}^{i,\mathcal{S}}_k\hat{\bs{x}}^i_{k|k-1} \right).
\end{equation}
In addition, the KCF introduces an explicit state-consensus correction to reduce disagreement among neighboring predicted estimates:
\begin{equation}
\label{eq:kcf_state_corr}
\Delta\hat{\bs{x}}^{i,\mathrm{state}}_k = \gamma_i\bm{P}^i_{k|k-1}
\sum_{j\in\mathcal{N}_i} \left(
\hat{\bs{x}}^j_{k|k-1} -
\hat{\bs{x}}^i_{k|k-1} \right),
\end{equation}
where $
\gamma_i =
\frac{\epsilon_{\mathrm{kcf}}}{1+\|\bm{P}^i_{k|k-1}\|_F}$,
$\|\cdot\|_F$ denotes the Frobenius norm, and $\epsilon_{\mathrm{kcf}}$ is a scalar tuning parameter.

The posterior state estimate is finally obtained as
\begin{equation}
\label{eq:kcf_final_state}
\hat{\bs{x}}^i_k
=
\hat{\bs{x}}^i_{k|k-1}
+
\Delta\hat{\bs{x}}^{i,\mathrm{meas}}_k
+
\Delta\hat{\bs{x}}^{i,\mathrm{state}}_k.
\end{equation}

The KCF is theoretically sub-optimal as it neglects the cross-covariances between local estimates that arise from the exchange of overlapping neighborhood information. Nevertheless, under the assumptions considered in~\cite{olfati2009kalman}, stability and asymptotic agreement of the local state estimates have been demonstrated for sufficiently small $\gamma$. The explicit state-consensus correction term therefore provides a scalable mechanism to reduce disagreement among neighboring predicted states, making the KCF a robust cooperative filtering strategy.

\subsection{Centralized Extended Kalman Filter (CEKF)}
A Centralized Extended Kalman Filter (CEKF) is also implemented as a reference solution. In this case, all the measurements obtained by the agents at time $t_k$ are processed by a single fusion node, and the information contributions are directly summed together.
\begin{equation}
\label{eq:cent_terms}
\bm{U}^{\mathrm{c}}_k=\sum_{i\in\mathcal{V}}\bm{U}^i_k,\qquad
\bs{\iota}^{\mathrm{c}}_k=\sum_{i\in\mathcal{V}}\bs{\iota}^i_k.
\end{equation}
The posterior centralized estimates can then be computed from the predicted ones as
\begin{equation}
\label{eq:cent_upd}
\begin{gathered}
\bm{P}^{\mathrm{c}}_k=\left[(\bm{P}^{\mathrm{c}}_{k|k-1})^{-1}+\bm{U}^{\mathrm{c}}_k\right]^{-1},
 \\
\hat{\bs{x}}^{\mathrm{c}}_k=
\bm{P}^{\mathrm{c}}_k\left[(\bm{P}^{\mathrm{c}}_{k|k-1})^{-1}\hat{\bs{x}}^{\mathrm{c}}_{k|k-1}+\bs{\iota}^{\mathrm{c}}_k\right].
\end{gathered}
\end{equation}
The prediction step instead follows the exact same approach described in \ref{sec:predict_step}.

\section{Numerical Simulations}
\label{results}
\subsection{Reference Scenario}
\begin{figure}
    \centering   \includegraphics[width=1.0\linewidth]{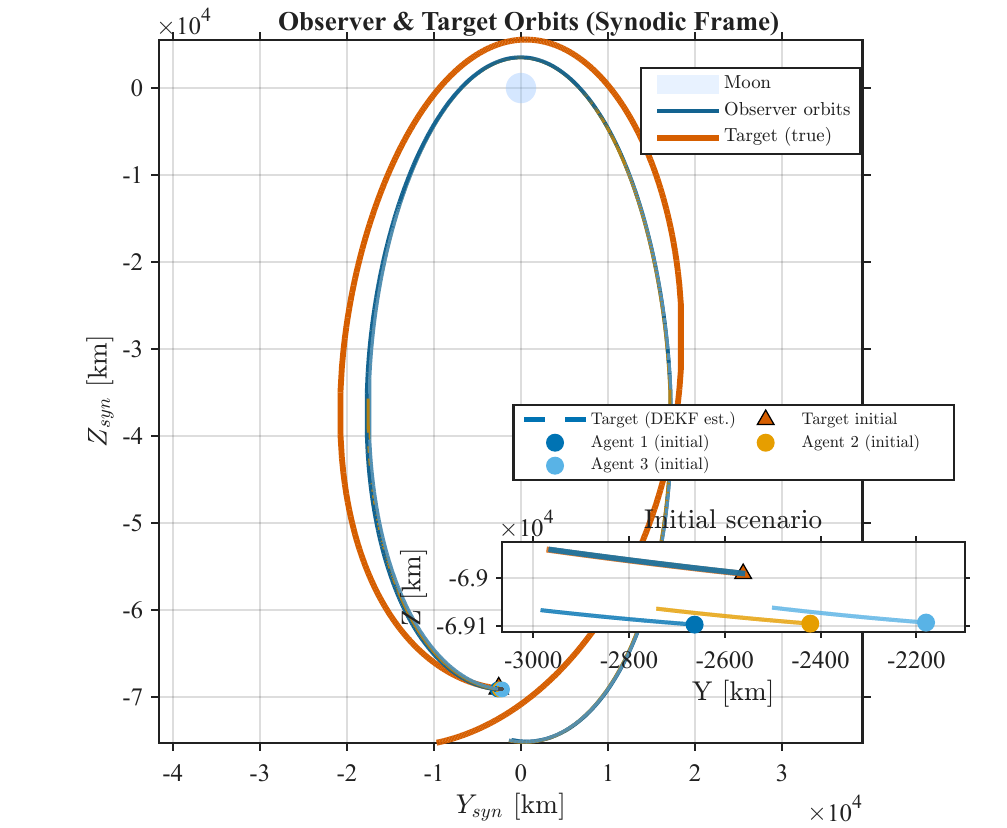}
    \caption{Reference scenario in synodic coordinates. The inset shows the initial condition for the observer formation and the target.}
    \label{fig:orbit}
\end{figure}

The filtering algorithms introduced in sec. \ref{sec:filters} were tested in the cislunar domain with three observer spacecraft tracking one target. Observer positions are assumed exactly known.
In particular, the reference orbit considered for the observer formation is a Southern Near-Rectilinear Halo Orbit (NRHO) around the L2 Lagrangian point of the Earth-Moon system (see Fig.~\ref{fig:orbit}). 
The observer satellites are initialized in the same NRHO orbit with different initial phases, resulting in an in-track formation where the spacecraft maintain a bounded relative separation. The target trajectory is generated by perturbing both the initial position and velocity with respect to the centroid of the observer formation.

The communication topology of the formation can be represented by the following undirected line graph

\begin{equation}
\label{eq:comm}
\begin{tikzpicture}[baseline=-0.5ex,node distance=1.8cm]
    \node[circle,fill=black,inner sep=2.2pt,label=below:{Agent 1}] (a1) {};
    \node[circle,fill=black,inner sep=2.2pt,label=below:{Agent 2},right of=a1] (a2) {};
    \node[circle,fill=black,inner sep=2.2pt,label=below:{Agent 3},right of=a2] (a3) {};
    \draw[-,line width=0.8pt] (a1) -- (a2);
    \draw[-,line width=0.8pt] (a2) -- (a3);
\end{tikzpicture}
\end{equation}

This results in a not-fully connected topology that allows the filtering algorithms to be tested under limited inter-agent communication while preserving the connectivity required for average consensus.

\subsection{Simulation Setup}
\label{sec:setup}
Each observer obtains a LOS measurement of the target every $\Delta t = 0.1$ min, only when the relative distance is below its visibility threshold of 500 km. The analyzed interval ensures simultaneous visibility from all three observers.  The angular noise standard deviation is set to
$\sigma_\theta = 10^{-4}$ rad, representative of onboard optical sensing,
including non-dedicated cameras~\cite{palo2013agile}. All filters are initialized with the same set of initial estimates $(\hat{\bs{x}}_0^i,\bm{P}_0^i)$, obtained from the geometry-based initialization scheme described in \cite{11520155}. The process noise covariance 
is set as
\begin{equation}
\label{eq:Q}
\bm{Q}=\mathrm{diag}(10^{-7},10^{-7},10^{-7},10^{-9},10^{-9},10^{-9}), 
\end{equation}
where the first three diagonal entries have units of $\mathrm{km}^2$ and the last three of $\mathrm{km}^2/\mathrm{s}^2$, per update. The measurement covariance matrix is computed from~\eqref{eq:R_reg} using the current LOS measurement.
In the nominal simulation scenario, the number of average consensus iterations used by the ICF and DEKF is set to $N_c=20$. The tuning parameter used to compute the KCF state-consensus gain $\gamma_i$ is chosen as $\epsilon_{\mathrm{kcf}}=0.1$ and the regularization coefficient is $\epsilon_{\mathrm{reg}} = 10^{-6}$.

\subsection{Results and Discussion}

\subsubsection{Nominal Scenario}
\begin{figure*}[!t]
    \centering
\includegraphics[width=0.8\linewidth]{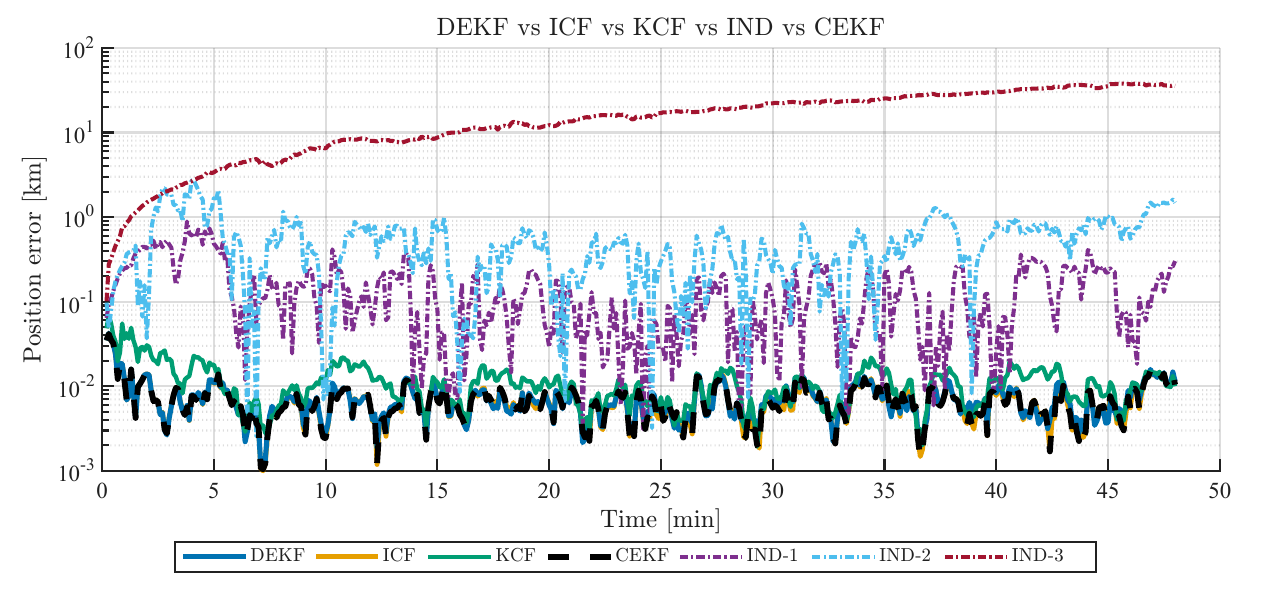}
    \caption{Comparison of the mean position error obtained by different filtering strategies in the nominal simulation case.}  
\label{fig:nominal}
\end{figure*}

 Fig.~\ref{fig:nominal} reports the Euclidean position error $\|\hat{\bs{r}}_t-\bs{r}_t^{\mathrm{true}}\|_2$, averaged over the three agents for each distributed filter. The CEKF error is shown separately, as are the individual independent-filter errors. The independent local filters are included in the comparison to highlight the benefit of inter-agent information exchange: each
agent is initialized identically to the distributed filters, but
performs the correction step using only its own local LOS
measurement, without exchanging information with neighboring
agents. 

The results show that the ICF and DEKF closely approach centralized performance over the considered visibility window. The KCF also constitutes a clear improvement with respect to the independent local filters; however, its error remains slightly higher and less aligned with the CEKF than those of the ICF and DEKF. In contrast, the independent local filters exhibit significantly larger errors, confirming that local LOS measurements alone are not sufficient to maintain accurate target-state estimates over the full window. 

\subsubsection{Sensitivity to the Number of Consensus Iterations}
To assess the sensitivity of the distributed filters to finite
communication, a Monte Carlo analysis is performed by varying the
number of average-consensus iterations \(N_c\). For each value of
\(N_c\), 100 Monte Carlo simulations are conducted with aligned
initial estimates. Fig.~\ref{fig:cons} reports the mean position
error averaged over the visibility window.

The DEKF exhibits a strong dependence on \(N_c\). For small values
of \(N_c\), the consensus stage provides only a poor approximation
of the network average, leading to degraded performance. As the
number of iterations increases, the DEKF error decreases
significantly and, for \(N_c \simeq 15\), approaches the CEKF
reference, consistently with the centralized-equivalence
property of the DEKF under ideal consensus and aligned priors.
The ICF shows a much weaker dependence on \(N_c\): even with few
consensus iterations, it remains close to the centralized reference,
whereas the DEKF requires a sufficiently converged consensus stage to
approach the same accuracy. The KCF, as expected, is insensitive to \(N_c\), since it relies on one-hop neighborhood fusion  rather than on iterative average consensus.
The ICF outperforms the KCF even for small \(N_c\). The remaining KCF gap may reflect both one-hop measurement availability and neglected cross-covariances; their individual effects are not isolated by this comparison.

\subsubsection{Sensitivity to Misaligned Initial Estimates}

In the second Monte Carlo analysis,  the number of consensus iterations is kept fixed to $N_c=20$, while different levels of initial prior mismatch are introduced among the agents. Starting from aligned initial estimates, each local prior is independently perturbed as
\begin{equation}
\label{eq:prior_mismatch}
\hat{\bs{x}}_{0,\alpha}^i=\hat{\bs{x}}_0^i+\alpha\bm{L}_0^i\bs{z}^i,
\end{equation}
where $\bm{P}_0^i=\bm{L}_0^i(\bm{L}_0^i)^\top$, $\bs{z}^i\sim\mathcal{N}(\bs{0},\bm{I}_6)$, and $\alpha$ is a scalar factor that controls the mismatch magnitude. For each value of $\alpha$, 100 random simulations are performed and the mean position error is computed over the visibility window.

\begin{figure*}[!t]
    \centering
    \begin{minipage}{0.48\linewidth}
        \centering
        \includegraphics[width=\linewidth]{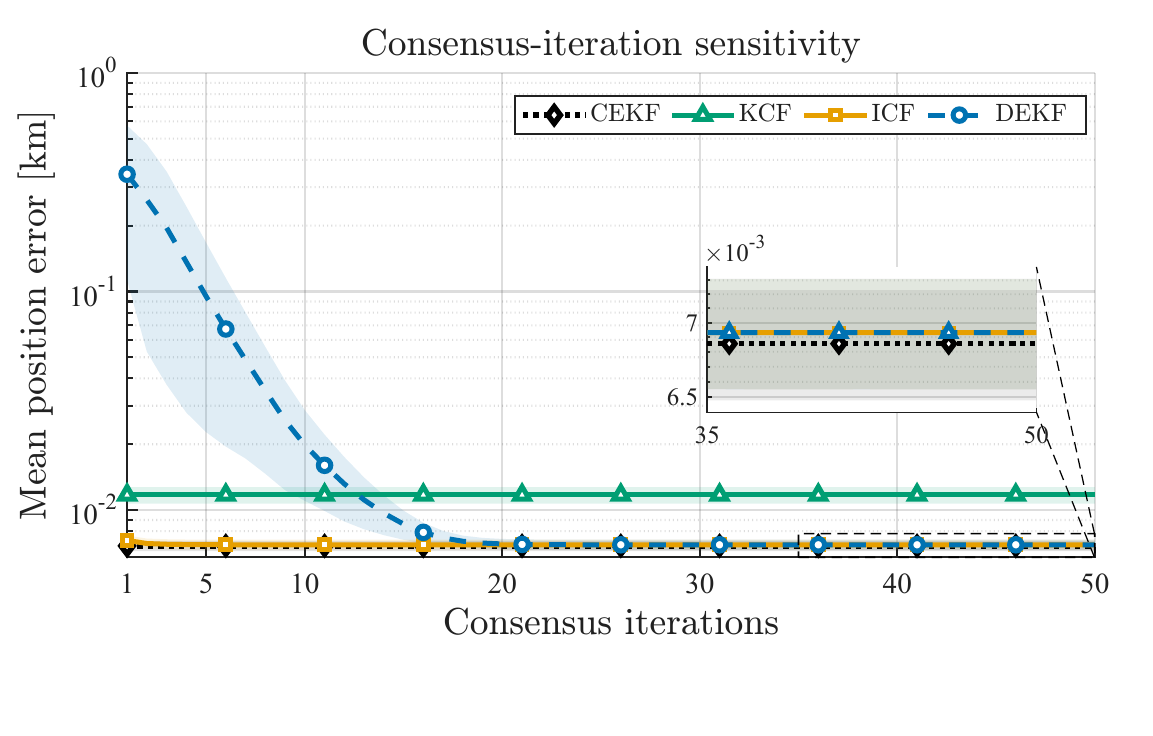}
        \caption{Monte Carlo sensitivity analysis with varying values of $N_c$. Solid lines and shaded regions denote the mean and $\pm1$ sample standard deviation across 100 Monte Carlo trials.}
        \label{fig:cons}
    \end{minipage}
    \hfill
    \begin{minipage}{0.48\linewidth}
        \centering
        \includegraphics[width=\linewidth]{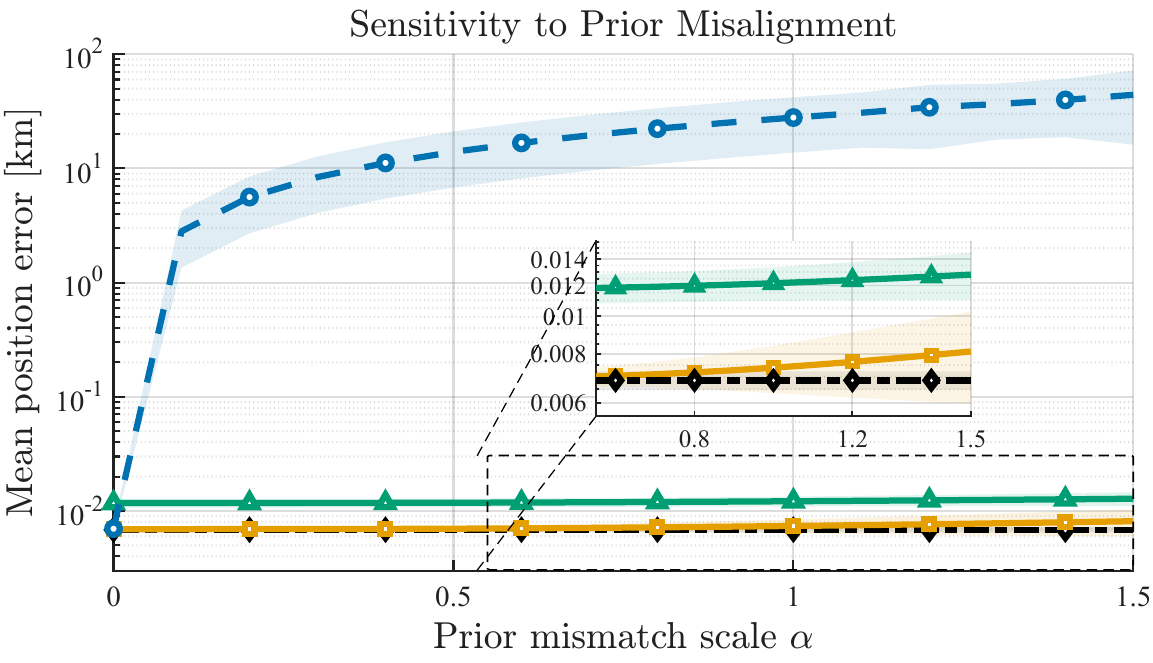}
        \caption{Monte Carlo sensitivity analysis with increasing values of $\alpha$. Solid lines and shaded regions denote the mean and $\pm1$ sample standard deviation across 100 Monte Carlo trials. Curve styles are as in Fig.~\ref{fig:cons}.}
        \label{fig:prior_mis}
    \end{minipage}
\end{figure*}
The results in Fig.~\ref{fig:prior_mis} show that the DEKF is highly sensitive to misaligned initial estimates. As the mismatch factor increases, the DEKF error grows rapidly, confirming that consensus on measurement information alone is not sufficient to compensate for discrepancies in the agents' priors. 
In contrast, the ICF and KCF exhibit a much weaker dependence on the prior mismatch than the DEKF. The ICF remains the closest distributed solution to the centralized reference over the considered range of $\alpha$, with only a moderate increase in the error as the prior mismatch grows. The KCF also remains bounded and significantly more robust than the DEKF, but it shows a larger residual gap with respect to the CEKF. 

\section{Conclusions}
\label{sec:conc}

This work investigated distributed filtering strategies for target tracking
in cislunar space using only angular measurements. Among the considered
approaches, the ICF provided the most reliable distributed approximation of
the centralized reference, remaining close to the CEKF both under finite
consensus iterations and increasing prior misalignment. The DEKF achieved
comparable accuracy when a sufficient number of consensus iterations was
available and the agents shared aligned prior estimates, but it showed a
strong sensitivity to limited communication and prior inconsistencies. The KCF offered a stable and communication-efficient cooperative alternative, significantly improving over independent local filters, although with a larger residual error with respect to the centralized reference. 

\bibliographystyle{IEEEtran}
\bibliography{bib}

\end{document}